# Seebeck Nanoantennas for Infrared Detection and Energy Harvesting Applications

Edgar Briones[1,*], Joel Briones[2], J. C. Martinez-Anton[3], Alexander Cuadrado[3], Stefan McMurtry[4], Michel Hehn[4], François Montaigne[4], Javier Alda[3], Javier González[1]

[1] CIACyT, Universidad Autonoma de San Luis Potosi, San Luis Potosi, 78210 SLP, Mexico
[2] Department of Mathematics and Physics, ITESO, Jesuit University of Guadalajara, 45604, Mexico
[3] Faculty of Optics and Optometry, Universidad Complutense de Madrid, 28037, Madrid, Spain
[4] Institut Jean Lamour, CNRS, Université de Lorraine, F-54506 Vandoeuvre Les Nancy, France

* Corresponding author: edgar.briones@uaslp.mx

*Abstract*— In this letter we introduce a new type of infrared sensor, based on thermocouple nanoantennas, which enables the energy detection and gathering in the mid-infrared region. The proposed detector combines the Seebeck effect, as a transduction mechanism, with the functionalities of the optical antennas for optical sensing. By using finite-element numerical simulations we evaluate the performance and optical-to-electrical conversion efficiency of the proposed device, unveiling its potential for optical sensing and energy harvesting applications.

## I. INTRODUCTION

Nanoantennas are metallic resonant structures at the core of new advances in photonics due to their capability to confine and manipulate light into a sub-wavelength scale [1,2]. These types of nanostructures take advantage of the wave-nature of radiation in order to induce a resonant current along their structure [3,4], which is subsequently used to sense or retrieve the optical energy [5,7].

Early works on infrared detection with optical antennas were performed by using niobium coupled-microbolometers in order to sense the antenna's resonant current [8,9]. These devices were successfully incorporated into large phase-arrays of optical antennas, leading to the development of faster thermal imagining acquisition systems, with efficiencies around 0.01 % [10]. Meanwhile, the use of coupled nano-rectifiers has enabled the nanoantennas to retrieve optical energy [11]. These so-called rectifying antennas (rectennas) seems to have a great potential in the field of energy harvesting since they exhibit a high theoretical efficiency (claimed to be 100 % [6,7]) and the ability to resonate at any-wavelength. In this context, several devices based on nanoantennas coupled to metal-insulator-metal (MIM), metal-insulator-insulator-metal (MIIM) and Esaki tunnel barriers, have been experimentally realized and measured as a proof-of-concept [12-17]. However, in spite of their attractive functionalities, actual rectifying nanoantennas have presented up to now a low efficiency, which is around $10^{-9}$ % [7,8,17]. In order to incorporate nanoantennas into harvesting applications different retrieving mechanisms must then be explored [18,19].

In this manuscript, we present a new type of infrared device based on the combination of nanoantennas for optical sensing with the Seebeck effect as a transduction mechanism [20-24]. These devices work by exploiting the temperature gradients, caused by the resonant current induced along their structure, when the nanoantennas are illuminated. The thermal gradients in turn generate a DC voltage $V_{OC}$ (by Seebeck effect) that can be sensed at the open edges of the structures, defining the retrieving energy mechanism [25]. This signal can be evaluated by:

$$V_{OC} = (S_A - S_B)\,\Delta T \qquad (1)$$

where $S_A$ and $S_B$ refers to the Seebeck coefficient of the metals that form the nano-thermocouple and $\Delta T$ refers to the temperature difference between the center and the open edges of the structure.

The proposed devices present some advantages when compared to the rectifying nanoantennas counterpart. Rectifying antennas have drawbacks on their responsivity and efficiency, which is due to the very different impedance between their elements (i.e. the high-speed rectifier and the nanoantenna), causing the efficiency to drop by a huge amount of several orders of magnitude. On the other hand, the performance of the current high-speed rectifiers is still low due to their poor diode-like behavior [17]. The incorporation of Seebeck nanoantennas can surpass those difficulties since it permits to discard the mismatch impedance (energy transfer) between elements. Seebeck nanoantennas are reappearing quite recently as good candidates to exhibit a better performance than antennas based on other transduction mechanisms [21-24].

## II. NUMERICAL SIMULATIONS

### A. Devices

The numerical analysis presented in this work was performed on a device consisting of a spiral bimetallic nanoantenna, shown in Fig. 1(a). The proposed geometry enables a broadband optical absorption leading to the confinement and enhancement of the incident optical field at the gap of the structures [26]. The nanoantenna lies on the top of a half-space $SiO_2$ substrate and its size was appropriately adjusted to resonate to mid-infrared wavelengths (40μm long).

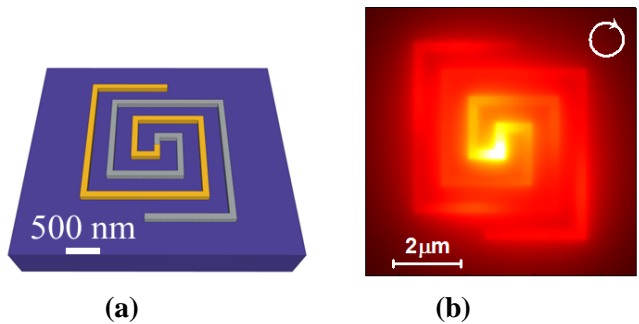

Fig. 1. (a) Schematic representation of the infrared Seebeck nanoantenna. The spiral nanoantenna is composed of two arms made-up of dissimilar metals joined at the center of the spiral. (b) Temperature map of the Seebeck nanoantenna 10.6 μm under circularly polarized illumination (taken from a plane 50nm below its surface).

The arms of the nanoantenna are made of titanium and nickel since those metals exhibit low-thermal conductivity ($\kappa_{Ni}$ = 90 W/mK and $\kappa_{Ti}$ = 21.9 W/mK), enhancing thus the thermal gradients along the spiral and the harvesting/detection of energy. These metals also show a considerable difference in their Seebeck coefficients ($S_{Ni}$ = 19.5 μV/K and $S_{Ti}$ = 7.19 μV/K [25]). The modeling of the device was performed by using COMSOL Multi-Physics ver3.5a (based on the finite-element method) commercial package that provides a good multi-physics platform where both the electromagnetic and the thermal domains are fully integrated [27]. The numerical model was built by using the reported optical and thermal properties of the materials, as an input into the solver, in the wavelength rage of interest (from 3 μm to 50 μm) [25,28].

The response of the Seebeck device is obtained at each single frequency by using a right-handed circularly polarized (RHCP) monochromatic plane-wave for normal far-field illumination. The irradiance $S$ of the wave was systematically adjusted for each frequency to be 117 W/cm² (intensity value currently used in antenna-based sensors measurements [9]).

*B. Thermal simulations and Seebeck voltage*

The map of temperature the device exhibits at 10.6μm is shown in Fig. 1(b). These numerical results were obtained by considering the nanoantenna as the only heat of source (Joule heating) and by solving the heat equation inside and outside the nanostructure, what unveils the distribution of the resonant current inside the structure [29]. From this map, the Seebeck voltage is easily derived from imaginary electrodes by using (1). From the color map it can be seen that the temperature increment $\Delta T$ between the gap and the extremes of the spirals is around 215 mK; this permits the antenna to act as a nano-thermocouple, whose output is around 5.7 μV. The Seebeck voltage for each single frequency is presented in Fig. 2. From the figure it can be appreciated the wide band of the device, which, as expected, is inherited from the optical properties of the antenna.

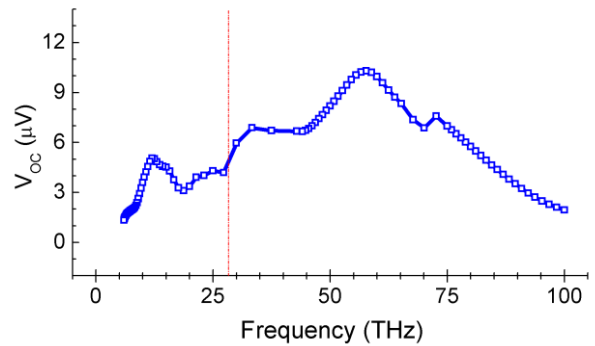

Fig. 2. DC voltage generated by the Seebeck nanoantenna under an irradiance of 117 W/cm², as a function of the frequency of the excitation.

### III. RESULTS

Fig. 3(a) shows the maximum DC power the proposed device can supply when incorporated into an electronic circuit. This amount of electrical energy is supplied only when the internal resistance of the acquisition system is tuned to match the DC resistance of the antenna $R_{int}$ (currently considered as the effective ohmic resistance of the materials [30,31]). By the other hand, the electronic circuit is expected to behave as a room-temperature thermal bath (taken as boundary conditions into the solver) keeping the outer ends of the arms as cool as possible, in order to maximize the thermal gradients inside the materials. When these two conditions met, the power supplied to the acquisition system is given by $P_{DC} = (V_{OC})^2/4R_{int}$ [31], where the internal resistance of the spiral geometry was estimated to be 490 Ω. The quoted absolute power was obtained by using a source of 117 W/cm². Even thought the used density flux is too high for (solar) harvesting applications, these results can be exploited to obtain the optical-to-electrical conversion efficiency of the device for every single frequency, expecting the efficiency to be independent from the intensity of the source. [25,28].

The optical-to-electrical conversion efficiency $\eta_e$ the device exhibits, at each single frequency, can be obtained by the ratio of the DC power $P_{DC}$ the antenna generates to the optical power that the antenna receives $P_{rec}$ as $\eta_e = P_{DC}/P_{rec}$. The optical power a nanoantenna collects $P_{rec} = A_{eff} \times S$ is found by using the reported collection area of the spiral nanoantenna $A_{eff}$ (12.5 μm² [9]) and the irradiance of the incidence beam S set for the simulations.

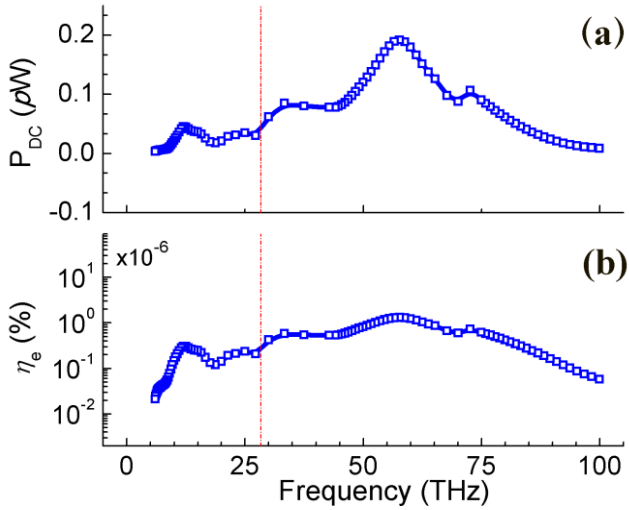

Fig. 3. Numerical results of the Seebeck nano antenna as function of the frequency of excitation. (a) Low-power DC generated by the device under an irradiance of 117 W/cm$^2$ and ideal electrical matching condition. (b) Percent optical-to-electrical conversion efficiency $\eta_e$(%).

The percent efficiency $\eta_e(\%) = \eta_e \times 100$ the Seebeck nanoantenna exhibits is shown in Fig. 3(b) as a function of the excitation frequency. The efficiency values range from $10^{-9}$ % to $10^{-5}$ %. These values are 1 to $10^3$ greater than the efficiencies reported for the rectifying antennas [17]. As we have previously mentioned, the performance of the rectifying antennas is drastically decreased by the unmatched impedance between the nano-rectifiers and the nanoantennas, as well as by the poor diode-like behavior of the tunnel barriers. By using Seebeck nanoantennas no impedance losses are seen, increasing this manner the nanoantennas overall performance. Moreover, the conversion efficiency could be increased by isolating the nanoantenna from the substrate in order to prevent the heat exchange between these two elements; by proceeding this way most of the optically induced heat will be exploited to induce the thermoelectric Seebeck voltage. From an experimentally point of view, this task could be achieved by suspending the device on air above its substrate (e.g., by using free-standing architecture).

## IV. CONCLUSIONS

In summary, the optical-to-electrical conversion efficiency of a Seebeck nanoantenna infrared detector was evaluated by performing numerical simulations (in the electromagnetic and thermal domain). The performed analysis shows that Seebeck nanoantennas represent an alternative technology to recover the free-propagating optical energy; increasing the overall performance by a $10^2$ factor when compared to the rectifying nanoantennas counterpart. Its performance can be increased by implementing technological strategies that could prevent energy losses by heat dissipation. Moreover, engineering of large phase-arrays of nanoantennas acting as series thermocouples arrays can be implemented to increase the performance of devices.

ACKNOWLEDGMENTS

Support from CONACYT-Mexico under postdoctoral grant CV-40859 (2015), Project ENE2009-013430 from the Spanish Ministerio de Innovación, Project "Centro Mexicano de Innovación en Energía Solar" from "Fondo Sectorial CONACYT Secretaría de Energía-Sustentabilidad Energética," and La Region Lorraine (France), is gratefully acknowledged..